\documentclass[prl,aps,twocolumn,showpacs]{revtex4}
\usepackage{amsmath,amssymb,graphicx,bm}
\usepackage{epsfig}
\usepackage{textcomp}

\newcommand{\qb}{{\bf q}}

\newcommand{\pb}{{\bf p}}
\newcommand{\xb}{{\bf x}}

\def\p12{p_{12}({\bf q},t)}
\begin{document}

\title{Minimizing the population extinction risk by migration}
\author{Michael Khasin$^{1,2}$,  Baruch Meerson$^3$, Evgeniy Khain$^2$, and Leonard M. Sander$^1$}
\affiliation{$^1$Department of
Physics, University of Michigan, Ann Arbor, MI 48109-1040, USA}
\affiliation{$^2$Department of Physics, Oakland University, Rochester, MI 48309, USA}
\affiliation{$^3$Racah Institute of Physics, Hebrew University of
Jerusalem, Jerusalem 91904, Israel}

\begin{abstract}
Many populations in nature are fragmented: they consist of local populations occupying separate patches. A local population is prone to extinction due to the shot noise of birth and death processes. A migrating population from another patch can dramatically delay the
extinction. What is the optimal migration rate that
minimizes the extinction risk of the whole population? Here we answer this question for a connected network of model habitat patches with different carrying capacities.

\end{abstract}
\maketitle

Many populations in nature are fragmented. Such \emph{meta-populations} consist of local populations occupying
separate habitat patches \cite{Levins,Hanski,Hanski2}. Habitat fragmentation is implicated in the decline and extinction of many endangered species \cite{endangered}.
To mitigate the negative impact of habitat fragmentation, conservation biologists have called for the construction of corridors to facilitate migration between separate habitat patches \cite{corridor}. Predicting how migration affects population persistence is important for species conservation, especially when the local population size
is depressed, and the local populations become prone to extinction because of randomness of the birth and death processes.  In this situation, it is of crucial importance to determine the optimal migration rate that maximizes the mean time to extinction (MTE) of the meta-population.   This problem has attracted much of attention from ecologists, and has been addressed, for different meta-populations, in  experiments  and stochastic simulations \cite{Ellner,Molofsky,Holoyak,Dey,Kerr,Shnerb}. Here we approach this important problem theoretically, for a simple logistic model of stochastic local populations coupled by migration. We analyze rare large fluctuations causing population extinction and show that there is an optimal migration rate that maximizes the MTE of the meta-population.

\textsc{Meta-population model}.  Mathematical biologists have proposed different types of stochastic meta-population models.
In a widely used class of models the local population distribution, its dynamics within a patch, and its effect on migration are ignored \cite{Gurney,McKane,Ross}. We show here that it is a proper account of these features that leads to the qualitatively new effect of the existence of an optimal migration rate.

Consider $N$ local populations of particles $A$ located on a connected network of patches $i=1,2,...,N$. The particles
undergo branching
$A\to 2A$ with rate constant $1$ on each patch and annihilation $2A\to \emptyset$ with rate constant  $1/(\kappa_i K)$ on patch $i$.  The parameters $\kappa_i={\cal O}(1)$, $i=1,2,\dots,N$, describe the disparity among the local carrying capacities $\kappa_i K$. Each particle can also migrate between connected patches $i$ and $j$ with rate constant $\mu_{ij}=\mu_{ji}$.  We assume that $\mu_{ij} = \mu M_{ij}$, where elements
of $M_{ij}$ are of order unity.

For $K\gg 1$  each local population is expected to be long-lived. Still, the shot noise will ultimately drive the whole meta-population to extinction. The MTE of the meta-population, $T$,  is exponentially large in $K$ but finite \cite{Shnerb,TREE}.
How does $T$ depend on the characteristic migration rate $\mu$? At $\mu=0$ each local population goes extinct separately, and $T_{\mu=0}$ is determined by the patch with the greatest carrying capacity, $K_m=K \max_i\{\kappa_i\}$:
\begin{equation}\label{muequal00}
   \ln T_{\mu=0}/K \simeq 2(1-\ln 2) \,\max_i\{\kappa_i\}
\end{equation}
($T$ for a single patch was found in Refs.  \cite{turner,ElgartPRE70,Kessler,AM1}). The ideas behind our results for $\mu>0$ are the following. At very fast migration, $\mu \to \infty$, the local populations become fully synchronized: both at the level of the expected local carrying capacities, and at the level of large fluctuations leading to population extinction. The total carrying capacity of the meta-population, as derived from the rate equation for this model \cite{suppl}, becomes $\bar{\kappa} K$, where
\begin{equation}\label{barkappa}
\bar{\kappa}=N^2 /\sum(\kappa_i^{-1}).
\end{equation}
One can argue, therefore, 
that at $\mu\to \infty$ the meta-population  goes extinct as if it were occupying a single effective patch with the total rescaled carrying capacity $\bar{\kappa}$, that is \cite{comparison}
\begin{equation}\label{fastgen}
\ln T_{\mu\to \infty}/K \simeq 2(1-\ln 2)\, \bar{\kappa}.
\end{equation}
The main result of our work is that, for unequal $\kappa_i$, $T$ reaches its maximum  at a finite value of the migration rate. This fact is intimately related to synchronization of the most probable local extinction events that occurs already at very small 
migration rates. The synchronization makes $T$ close to that for a single patch with the \emph{combined} carrying capacity $K \sum_i\kappa_i$:
\begin{eqnarray}
  &&\ln T_{\mu \to 0}/K \simeq  2 (1-\ln 2) \sum_i \kappa_i . \label{S1N}
\end{eqnarray}
Now let us inspect the MTE as described by Eqs.~(\ref{muequal00}), (\ref{fastgen}) and (\ref{S1N}).  As $\sum_i\kappa_i\geq\max_i\{\kappa_i \}$ and $\sum_i\kappa_i \ge \bar{\kappa}$ for any $\kappa_i$, the MTE must reach a maximum at a finite value $\mu=\mu_*$, unless all the patches have the same carrying capacity. We will present evidence that 
$\mu_*\ll 1$ and scales as $1/K$.

How to understand qualitatively the non-trivial dependence of the MTE on $\mu$? Consider first the large-$\mu$ regime. Equation~(\ref{barkappa}) implies that patches with \emph{smaller} carrying capacities dominate the effective annihilation rate. For example, in a system of two patches, each particle spends half its time on each of the two patches. Then the patch with the smaller carrying capacity dominates the total annihilation rate. As $\mu$ decreases, particles will spend enough time on the good patch so that the total carrying capacity will drift up, and the MTE will increase. Now consider a very small
but finite $\mu$, so that the migration rate is higher than the (exponentially small) local extinction rates. Here, for the whole meta-population to go extinct, all local extinction events must occur in synchrony, and this leads to Eq.~(\ref{S1N}).

Now we expose our results in more detail. For simplicity, we will first consider a system of two patches and then generalize our results to a network of $N$ patches. The rate equations for the two-patch system are:
\begin{eqnarray}
\label{eq:mf}
\dot x&=& x-x^2-\mu x+\mu y\,, \nonumber \\
\dot y&=&y-\frac{y^2}{\kappa}+\mu x-\mu y,
\end{eqnarray}
where $x$ and $y$ are the local population sizes rescaled by $\kappa_1 K$, and $\kappa=\kappa_2/\kappa_1$.  Equations (\ref{eq:mf}) have two fixed points: the unstable point $x_0=y_0=0$ that describes an empty system, and a stable point $[x_*(\kappa,\mu)>0,\, y_*(\kappa,\mu)>0]$ that describes an established meta-population.  
At $\mu= 0$ one has $x_*=1$ and $y_*=\kappa$, whereas for  infinitely fast migration, $\mu \to \infty$,
\begin{equation}\label{harmonicmean}
    x_*=y_*=2 \kappa/(1+\kappa).
\end{equation}
The characteristic time $t_r$ of population establishment is determined by the smaller of the two eigenvalues  of the linear stability matrix of Eqs.~(\ref{eq:mf}) at the fixed point $(x_*,y_*)$.

In a stochastic formulation, the probability $P_{m,n}(t)$ to find $m$ particles in patch 1 and $n$ particles in patch 2 evolves in time according to the master equation
\begin{eqnarray}
\label{eq:master_master}&&
\dot{P}_{m,n}(t)=\hat{H} P_{m,n}\equiv (m-1)P_{m-1,n}+(n-1)P_{m,n-1} \nonumber \\
&&+\frac{(m+1)(m+2)}{2 K}P_{m+2,n}+\frac{(n+1)(n+2)}{2 \kappa K}P_{m,n+2} \nonumber \\
&&+\mu (m+1)P_{m+1,n-1}+\mu (n+1)P_{m-1,n+1} \nonumber\\
&&-\!\!\left[(1+\mu)(m+n)+ \frac{m(m-1)}{2K}+\frac{n(n-1)}{2\kappa K}\right]\!\!P_{m,n}.
\end{eqnarray}
The probability $P_{0,0}$ that the meta-population
goes extinct by time $t$ is governed by the equation
\begin{equation}\label{probext}
    \dot{P}_{0,0}(t)=\frac{1}{K} P_{2,0}+\frac{1}{\kappa K} P_{0,2}.
\end{equation}
\textsc{Long-time dynamics and the MTE}. For $t\gtrsim t_r$,  $P_{m,n}(t)$ becomes sharply peaked at the local carrying capacities $m_*=Kx_*$ and $n_*=Ky_*$, corresponding to the stable fixed point $(x_*, y_*)$ of the mean-field theory. The subsequent slow decay of $P_{m,n}$ in time is determined by the lowest excited eigenmode $\pi_{m,n}$ of the master equation operator $\hat{H}$: $P_{m,n} (t) \simeq \pi_{m,n} \exp(-t/T)$. Simultaneously, a probability peak at $m=n=0$ grows with time: $P_{0,0}(t) \simeq 1- \exp(-t/T)$  \cite{AM2006,AssafPRE81,MS11a}.
The inverse eigenvalue $T$ is an accurate approximation to the MTE. Since it turns out to be exponentially large with respect to $K\gg 1$, one can
neglect the right-hand-side of the eigenvalue problem $\hat{H} \pi_{m,n}=\pi_{m,n}/T$ and consider
the quasi-stationary equation $\hat{H} \pi_{m,n}\simeq 0$. Once $\pi_{m,n}$ is found, the MTE can be determined from Eq.~(\ref{probext}):
\begin{equation}\label{tau}
    T= [\pi_{2,0}/K+\pi_{0,2}/(\kappa K)]^{-1}
\end{equation}
\textsc{WKB theory}. To find $\pi_{m,n}$ for not too small values of $\mu$, we employ a dissipative variant of Wentzel-Kramers-Brillouin (WKB) approximation, pioneered in Refs. \cite{Kubo,Hu,Peters,DykmanPRE100}, and extensively used in the problems of stochastic population extinction \cite{ElgartPRE70,Kessler,AssafPRE81,MS11a,DykmanPRL101,MeersonPRE77,KamMS,AKMmod,MeersonPRE79,
KhasinPRL103,MeersonPRE81,AMS,LM2011,GM}, see also Ref. \cite{Doering}. The WKB ansatz is
\begin{equation}\label{WKB}
    \pi_{m,n} = \exp[-KS(x,y)],
\end{equation}
where $x=m/K$ and $y=n/K$ are treated as continuous variables. We plug Eq.~(\ref{WKB}) into the quasi-stationary equation  $\hat{H} \pi_{m,n}= 0$ and Taylor expand $S$ around $(x,y)$. In  leading order in $1/K\ll 1$ this gives a zero-energy Hamilton-Jacobi equation $H(x, y, \partial_x S, \partial_y S) = 0$ with  classical Hamiltonian
\begin{eqnarray}
&&H(x,y,p_x,p_y) \!\!=\!\!x\left(e^{p_x}-1\right)+\frac{x^2}{2} \left(e^{-2p_x}-1\right)\nonumber \\
&&+y\left(e^{p_y}-1\right)+\frac{y^2}{2\kappa} \left(e^{-2p_y}-1\right) \nonumber \\
&&+\mu x\left(e^{-p_x+p_y}-1\right)+\mu y\left(e^{p_x-p_y}-1\right). \label{H}
\end{eqnarray}
The established population corresponds to the fixed point $M=(x_*, y_*, 0,0)$ of the Hamiltonian flow. Up to a pre-exponent, $T \sim \exp (K {\cal S})$,
where ${\cal S}$ is the action along the \emph{instanton}: a special zero-energy ($H=0$) trajectory  in the phase space $(x,y,p_x,p_y)$ that exits, at time $t=-\infty$, the fixed point $M$ and approaches the fluctuational extinction point $F$ that, for the two-patch branching-annihilation model, is $(0,0,-\infty,-\infty)$
\cite{infinity}.  In the absence of an independent integral of motion in addition to the Hamiltonian itself, this trajectory, and the action along it, can only be found numerically. Analytical results are possible in the limits of small and large $\mu$ that we will now consider.

When $\mu\to 0$
the Hamiltonian (\ref{H}) becomes separable, and the instanton trajectory can be easily found:
\begin{eqnarray}
   &&x(t) = q(t-\tau_x),\;\;\;y(t)=\kappa q (t-\tau_y)\nonumber\\
   && p_x(t)=p(t-\tau_x), \;\;\;  p_y(t)=p(t-\tau_y),\label{loc}
\end{eqnarray}
where
\begin{equation}\label{qp}
 q(t)=2 (2+3 e^{t}+e^{2t})^{-1},\;\; p(t)=-\ln (1+e^{t}).
\end{equation}
Notice that the solution for $\mu \to 0$ includes arbitrary time
shifts $\tau_x$ and $\tau_y$ in the $x$- and $y$ populations, respectively. These will become important shortly. The action
\begin{eqnarray}
  {\cal S}(\mu\to 0) &=& \int_{-\infty}^\infty \left( p_x \dot{x} + p_y \dot{y}- H\right)dt \nonumber \\
  &=& 2 (1-\ln 2) (1+\kappa) \simeq \ln T_{\mu\to 0}/K. \label{S1}
\end{eqnarray}
Equation~(\ref{S1}) coincides with that for an effective one-patch system with the \emph{combined} carrying capacity $(1+\kappa) K$. This extinction time is exponentially large compared with the one obtained if one neglects migration completely, see Eq.~(\ref{muequal00}) with $\max \{\kappa_i\}=1$. The sharp increase of $T$ once slow migration is allowed results from synchronization of the most probable local extinction paths (\ref{loc}). For $\mu \ll 1$, the two noisy  local populations behave almost independently for typical, small fluctuations. For rare large fluctuations, such as the one causing extinction of the whole meta-population, the dynamics of the local populations becomes synchronized. How does the synchronization show up in the WKB calculations? In the absence of migration, $\mu=0$, the time shifts $\tau_x$ and $\tau_y$ which appear in Eqs.~(\ref{loc}) are arbitrary, reflecting the time-translational invariance of local extinctions. A small $\mu>0$ partially breaks this invariance and selects a particular relative time shift $\tau=\tau_y-\tau_x$, implying synchronization. 
Since the zero-order action (\ref{S1}) is invariant with respect to the local time shifts, it is necessary to consider a small $\mu$ correction, ${\cal S}={\cal S}(\mu\to 0)+\Delta {\cal S}$ in order  to determine  $\tau$. The first order correction can be calculated by integrating over the unperturbed $x$- and $y$-instantons (\ref{loc}):
\begin{eqnarray}
  \Delta {\cal S} &=& -\mu \,\max_{\tau} \chi (\tau),\nonumber \\
  \chi(\tau)&=&\int_{-\infty}^{\infty} \left\{q(t)\left[e^{-p(t)+p(t-\tau)}-1\right]\right.\nonumber\\
   &+&\left. \kappa q(t-\tau)\left[e^{p(t)-p(t-\tau)}-1\right] \right\} dt. \label{linear}
\end{eqnarray}
That is, the optimal time shift $\tau=\tau_*(\kappa)$ is determined from the minimization of the action, that is the maximization of $\chi(\tau)$, with respect to $\tau$. This minimization can be easily performed, as the  integral in Eq.~(\ref{linear}) can be evaluated analytically \cite{suppl}.

\begin{figure}[ht]
\includegraphics[width=2.5 in,clip=]{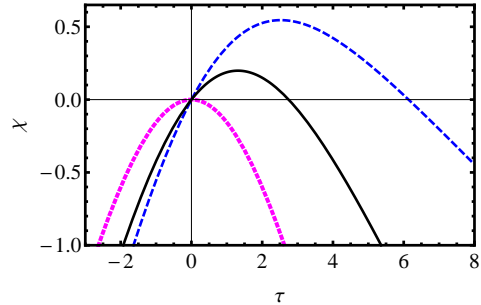}
\caption{(Color online) Function $\chi(\tau)$, see Eq.~(\ref{linear}), for two patches, for $\kappa=1$ (dotted line),  $0.5$ (solid line) and $0.25$ (dashed line).}
\label{chi}
\end{figure}

\begin{figure}[ht]
\includegraphics[width=3.0 in,clip=]{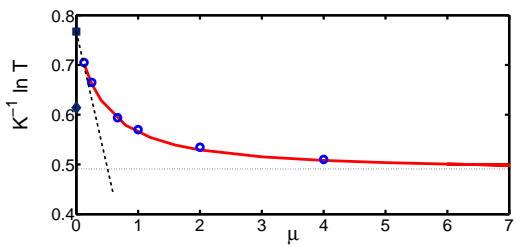}
\caption{(Color online) $K^{-1}\ln T$ vs. the migration rate $\mu$ for two patches, $\kappa=0.25$. Circles: numerical WKB solutions.  Diamond and square: predictions of Eqs.~(\ref{muequal00}) and ~(\ref{S1}), respectively. Dashed line: prediction of Eq.~(\ref{linear}) for $\mu\ll 1$: $\delta {\cal S}=-0.5455\dots \cdot \mu$. Dotted line: prediction of Eq.~(\ref{fast}) for $\mu\gg 1$. The solid line was obtained from a numerical solution of the master equation (\ref{eq:master_master}) for $K=220$.}
\label{II}
\end{figure}

By virtue of Eq.~(\ref{loc}), $\chi(0)=0$. This implies that $\Delta {\cal S}\le 0$, and so $T$ is a non-increasing function of $\mu$ for $\mu \ll 1$. The function $\chi(\tau)$ is depicted in Fig. \ref{chi} for $\kappa=1$, $0.5$ and $0.25$. For $\kappa=1$ (two identical patches)
the maximum is achieved at $\tau=0$, as expected from symmetry, so $\Delta S = 0$. In this case the solution (\ref{loc}) with $\kappa=1$ holds for \emph{all} $\mu$. That is, a higher migration rate does not affect $T$ up to a pre-exponential factor. For $\kappa<1$  we obtain  $\tau_*(\kappa)>0$ and $\Delta {\cal S}<0$, that is $T$ goes down with an increase of $\mu$, see Fig. \ref{II}. Because of the large factor $K$, a small decrease in ${\cal S}$ translates into an exponentially large reduction of $T$ of the meta-population. Note that the WKB approximation, leading to Eq.~(\ref{linear}), is only valid for $\mu \gg K^{-1}$. We expect that,  for $\mu \lesssim 1/K$
(but not exponentially small in $K$), \emph{weak} synchronization [to within time uncertainty of $(\mu K)^{-1}$] occurs,
again leading to MTE as in Eq.~(\ref{S1}) \cite{degenerate}.

Now consider the opposite limit, $\mu \to \infty$. Here the total population size  $Q=x+y$  varies  slowly in comparison with the fast migration. The fast variables
$x$ and $y$ rapidly adjust to the slow dynamics of $Q$, staying close to their stationary values for the
instantaneous value of $Q$. Transforming to $Q$ and $q=x$ and associated conjugate momenta as a new set of canonical variables, one arrives \cite{suppl} at a Hamiltonian, associated with the large fluctuations  of the total population size:
\begin{eqnarray}
  &&H_{slow}(Q,P) = \varepsilon  H_1[\tilde{q}(Q),Q,0,P] \nonumber \\
   &&=\varepsilon \left[Q (e^{P}-1)+\frac{1+\kappa}{8 \kappa} Q^2\left(e^{-2 P}-1\right)\right],
   \label{slow}
\end{eqnarray}
with $\varepsilon=1/\mu \ll 1$. Equation~(\ref{slow}) describes an effective single-patch Hamiltonian with a rescaled carrying capacity $\bar{\kappa}=4 \kappa/(1+\kappa)$, and we obtain
\begin{equation}\label{fast}
   \frac{\ln T_{\mu \to \infty}}{K} = \frac{8 (1-\ln 2) \kappa}{1+\kappa}.
\end{equation}
For $N=2$ this agrees with the announced result~(\ref{fastgen}). 

\textsc{WKB numerics}. For intermediate values of $\mu$ the instantons, and the associated action, can be found numerically: either
by a shooting method \cite{MeersonPRE77,DykmanPRL101}, or by 
iterations \cite{ElgartPRE70,LM2011,Stepanov}. Here we used both methods, and the results for $\ln T/K$ agreed within  less than $1$ per cent.  Figure~\ref{II} shows the numerically found ${\cal S}$ for $\kappa=0.25$ and different $\mu$, respectively. At $\mu\ll 1$ the numerical results agree with the prediction of linear theory, Eq.~(\ref{linear}). At large $\mu$ they approach the asymptote (\ref{fast}). Similar results were obtained for other values of $\kappa$. Figure~\ref{II} also compares the WKB results with those of a numerical solution of (a truncated version of) the full master equation
(\ref{eq:master_master}).  

\begin{figure}[ht]
\includegraphics[width=3.5 in,clip=]{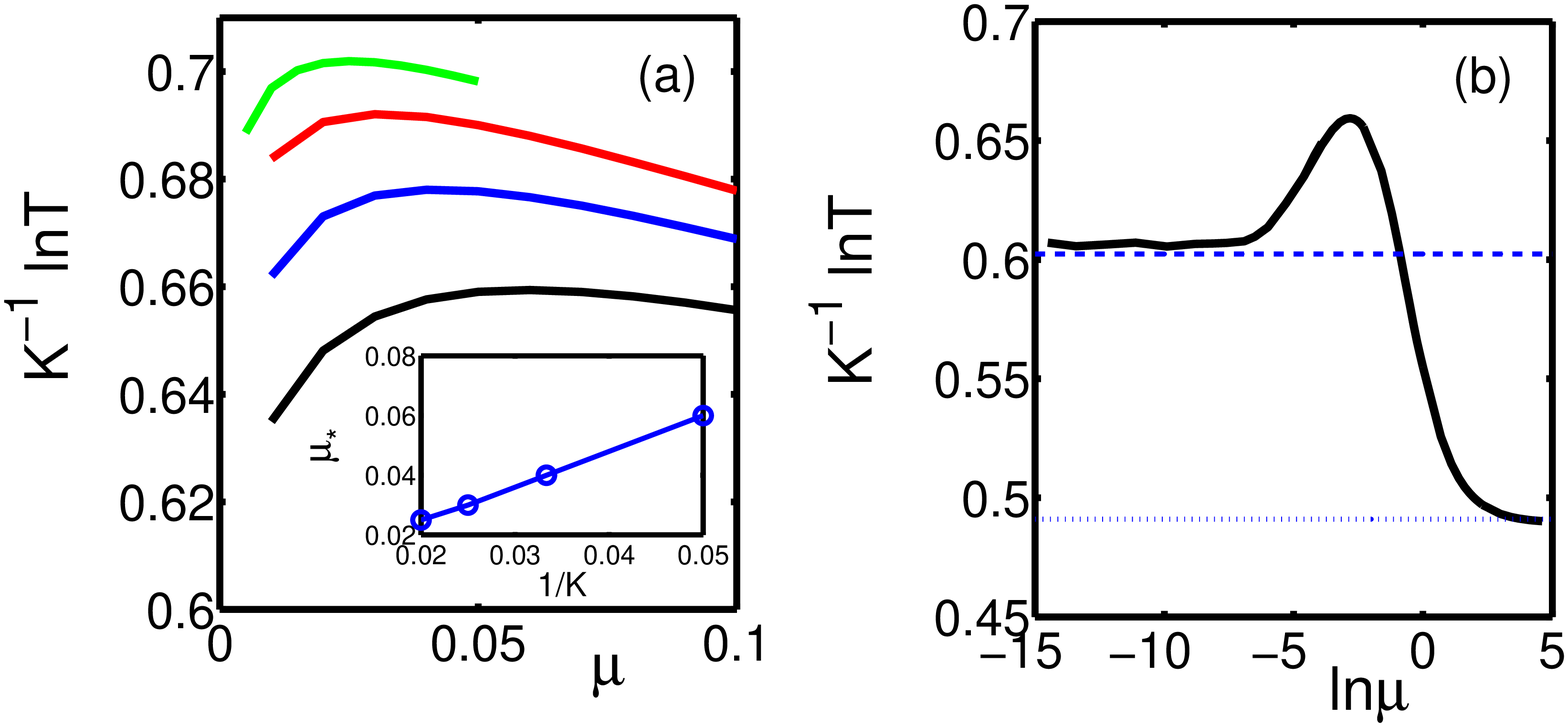}
\caption{(Color online) $K^{-1}\ln T$ vs. $\mu$ (a) and $\ln \mu$ (b)
for a two-patches meta-population from a numerical solution of the master equation and stochastic simulations. (a) $\kappa=0.25$ and $K=20, 30, 40$ and $50$ (bottom to top).  Inset: the migration rate $\mu_{*}$, at which the maximum of MTE is observed, versus $K$. (b) $\kappa = 0.25$ and $K = 20$; dashed line: Eq.~(\ref{muequal00}),
dotted line: Eq.~(\ref{fast}).}
\label{III}
\end{figure}

\textsc{Beyond  WKB theory}. To evaluate the maximum MTE and the optimal migration rate, one needs to resolve the jump of $(\ln T)/K$ at $\mu=0$ predicted by the WKB theory, see Eqs.~(\ref{muequal00}) and~(\ref{S1}). We determined the MTE for exponentially small $\mu$ by  numerically solving the master equation (\ref{eq:master_master}) and by performing stochastic simulations.
The resulting $\mu$-dependence of the MTE, at $\kappa=0.25$ and different $K$, is shown in Fig.~\ref{III}. The maximum of $T$ is observed at a small migration rate $\mu_*$ that apparently scales as $K^{-1}$.

\textsc{Network of $N$ patches.} Our results can be generalized to a connected network of $N$ patches with  migration rate $\mu_{ij}$ between any two connected patches $i$ and $j$. We assume $\mu_{ij}=\mu M_{ij}$, where $M_{ij}=M_{ji}\sim 1$.  For $\mu=0$, $T$ is given by Eq.~(\ref{muequal00}). For $\mu \to \infty$, the rescaled population size of each patch becomes $x_i=Q/N$, where the total size of the meta-population $Q=\sum_i x_i$ slowly varies in time. We can perform a canonical transformation from $x_N$ to the new coordinate $Q$, keeping $x_1, x_2,\dots, x_{N-1}$ the same. This procedure \cite{suppl} brings about a slow one-population Hamiltonian $H_{slow}(Q,P)$, with $P=p_{x_N}$ and effective carrying capacity $\bar{\kappa}$ from Eq.~(\ref{barkappa}):
\begin{eqnarray}
  &&H_{slow}=\varepsilon \left[Q (e^{P}-1)+ \frac{1}{2 \bar{\kappa}}Q^2\left(e^{-2 P}-1\right)\right].
\end{eqnarray}
This immediately yields  the announced result (\ref{fastgen}). 

How does synchronization of the most probable local extinction paths for small $\mu$ work for the $N$ patches? For $\mu \to 0$ the WKB-instanton is described by $x_i(t)=\kappa_i\, q(t-\tau_i)$, $p_i(t)=p(t-\tau_i)$, where
$\tau_i$ are constants, $i=1,2,...,N$, and functions $q(t)$ and $p(t)$ are defined in Eq.~(\ref{qp}).
This leads to the
action
\begin{eqnarray}
  &&{\cal S}(\mu\to 0) =  2 (1-\ln 2) \sum_i \kappa_i \simeq \ln T_{\mu \to 0}/K,\label{S1N1}
\end{eqnarray}
as announced in Eq.~(\ref{S1N}). The relative time shifts $\tau_i$, $i=1,2,\dots,N-1$, determining synchronization of the local extinction paths, can be found similarly to Eq.~(\ref{linear}), by minimizing 
$\Delta {\cal S}$:
\begin{eqnarray}
  &&\Delta {\cal S} = -\mu \,\max_{\{\tau_i\}}  \int_{-\infty}^{\infty} \sum_{i=1}^N \chi_i(t,\bm{\tau}) dt, \label{SN} \\
&&\chi_i(t,\bm{\tau})= \kappa_i q(t-\tau_i)\,\sum_{j\in {\cal{I}}_i}M_{ij} \left[e^{p(t-\tau_j)-p(t-\tau_i)}-1\right], \nonumber
\end{eqnarray}
where ${\cal{I}}_i$ is the subset of indices, corresponding to the patches directly connected to the patch $i$. As in many other problems with
multi-dimensional instantons \cite{book},
the minimization will typically give a unique solution up to the overall time shift. 
Since for the unperturbed local instantons $p_i(t)$ is independent of $i$,  we have $\chi_i(t,\mathbf{0})=0$  in Eq.~(\ref{SN}). Therefore,  $\Delta S \le 0$, i.e., 
$T$ is a non-increasing function of $\mu$, as in $N=2$ case. Generically, $\Delta S$ is strictly negative, so $T$ decreases with an increase of $\mu$ for small $\mu$. If all patches have the same carrying capacity, $T$ is constant in the WKB regime, up to a pre-exponential factor,  and corresponds to a single-patch MTE with the combined carrying capacity $K N$ \cite{ElgartPRE70,MS11a}.  Finally,  the validity of the WKB theory demands $\mu n\gg K^{-1}$, where $n$ is a typical number of connections of a patch (that is, a typical node degree of the network), i.e., $|{\cal I}_i|\sim n$, whereas the WKB perturbation theory for small $\mu$ demands $\mu n \ll 1$.

\textsc{In summary}, we have developed a quantitative theory of stochastic extinction of an established meta-population where individuals can migrate between different habitat patches. We have found that, as the meta-population goes extinct, local extinction paths become synchronized already at very small migration rates. 
Finally, we have shown that the MTE of the meta-population reaches its maximum for a small but nonzero migration rate.
B.M. was supported by the Israel Science Foundation (Grant No. 408/08),
by the US-Israel Binational Science Foundation (Grant No. 2008075), and by the Michigan Center for Theoretical Physics.

\begin{large}
\begin{center}
\textbf{Appendix}
\end{center}
\end{large}
\renewcommand{\theequation}{A\arabic{equation}}
\setcounter{equation}{0}
We present here derivations of several results outlined in the main text.
\subsection{\textit{Effective carrying capacity of the meta-population in the fast-migration limit}}
Here we provide  more detail on the deterministic dynamics in the fast-migration limit.
The rate equations for $N$ patches with local carrying capacities $K_i=K \kappa_i$ are
the following:
\begin{eqnarray}
\label{eq:mf1}
\dot x_i&=& x_i-x_i^2/\kappa_i+\mu \sum_{j\in {\cal I}_i}M_{ij}\left(x_j-x_i\right),\,
\end{eqnarray}
for $i=1,\dots,N$. Here ${\cal I}_i$ is the set of indices of patches connected to patch $i$, and
$M_{ij}=M_{ji}$.  For $\mu \rightarrow \infty$ we obtain
\begin{eqnarray}
\label{eq:mfb}
\sum_{j\in {\cal I}_i}M_{ij}\left(x_j-x_i\right)=0,\;\;\; i=1,\dots,N.
\end{eqnarray}
Equations~(\ref{eq:mfb}) are solved by $x_i=x_j=x$ for all $i$ and $j$. Now, summing up Eqs.~(\ref{eq:mf1}) and replacing $x_i$ by $X/N=x$, where $X=\sum_1^N x_i$ is the total meta-population size, we obtain:
\begin{eqnarray}
\label{eq:mf2}
&&\dot X= X- X^2/\bar{\kappa} \,, \nonumber \\
&&\bar{\kappa}=N^2 /\sum(\kappa_i^{-1}).
\end{eqnarray}
Therefore, in the fast-migration limit, the total meta-population size evolves as if the population were occupying a single patch with effective carrying capacity $\bar{K}\equiv K \bar{\kappa}$, as
announced in Eq.~(2) of the main text.

\subsection{\textit{Slow-migration limit in WKB regime}}
\subsubsection{General case: N patches}
To calculate the correction to WKB action in the first order in $\mu$, described by Eq.~(20) of the main text,
we need to evaluate the following integral:
\begin{eqnarray}
&&I(\tau )=  \int_{-\infty}^{\infty} q(t)\left[e^{p(t-\tau )-p(t)}-1\right] dt,\label{SN2}
\end{eqnarray}
so that
\begin{eqnarray}
 \!\!\!\! &&\Delta {\cal S} = -\mu \,\max_{\{\tau_i\}} \sum_{i, j\in {\cal{I}}_i} \kappa_i M_{ij} I\left(\tau_{j}-\tau_i\right). \nonumber
\end{eqnarray}
The integral $I(\tau)$ can be evaluated analytically, using the explicit one-patch instanton solution, see Eq. (13) of the main
text.
Upon the change of variables $\xi=e^{t}$, we obtain
\begin{eqnarray}
I(\tau )&=& 2(e^{\tau}-1) \int_{0}^{\infty} \frac{d\xi}{(2+3 \xi+\xi^2)(e^\tau+\xi)}  \nonumber \\
&=& \frac{2 \left[ \left(e^{\tau}-1\right)\ln 2-\tau\right]}{e^{ \tau}- 2}.
\end{eqnarray}
The first-order correction to the action becomes, therefore,
\begin{eqnarray}
 \!\!\!\!\!\! \!\!\!\!\!\! \!\!\!\Delta {\cal S}\!=\! \mu \,\min_{\{\tau_i\}}  \!\sum_{i, j\in {\cal{I}}_i}  \frac{ 2\kappa_i M_{ij}\left[ \left(e^{\tau_j-\tau_i}-1\right)\ln 2-\tau_j+\tau_i\right]}{2-e^{ \tau_j-\tau_i}}. \label{SN3}
\end{eqnarray}
The minimization of this sum with respect to the set of time differences $\tau_i-\tau_j$ will generically give a unique solution, up to an arbitrary overall time shift $\tau_0$. 

\subsubsection{Example: Two patches}
For $N=2$ we can put $\kappa_1=1$, $\kappa_2=\kappa$ and $M_{12}=1$.  Setting, without loss of generality, $\tau_0=0$,  we obtain from Eq.~(\ref{SN3}):
\begin{eqnarray}
\Delta {\cal S}&=& 2\mu\,\min_{\tau}  \left[\frac{\left(e^{\tau}-1\right)\ln 2-\tau}{2-e^{ \tau}}\right.\nonumber \\
&+& \left.\kappa  \frac{\left(e^{-\tau}-1\right)\ln 2+\tau}{2-e^{ -\tau}}\right]. \label{SN4}
\end{eqnarray}
The minimization of this expression can be easily performed numerically. For $\kappa=0.25$, one obtains $\Delta {\cal S}= -0.5455 \dots \mu$. This asymptotic is shown  as the dashed line in  Fig. 2 of the main text.

\subsection{Fast-migration limit in WKB regime}
In the limit of large $\mu$ it is convenient to rescale time by $\mu$ and present the Hamiltonian  as $H=H_0+ \varepsilon H_1$ with $\varepsilon=1/\mu \ll1$:
\begin{eqnarray}
\label{eq:ham}
&&H_0\!=\!\frac{1}{2}\sum_{i} \sum_{j\in {\cal I}_i}M_{ij}\left[ x_i\left(e^{p_j-p_i}-1\right)+ x_j\left(e^{p_i-p_j}-1\right)\right], \nonumber \\
&&H_1=\sum_{i=1}^N x_i\left(e^{p_i}-1\right)+\frac{x_i^2}{2\kappa_i} \left(e^{-2p_i}-1\right).\nonumber
\end{eqnarray}
It is the ``migration Hamiltonian" $H_0$ that dominates the dynamics in this limit, whereas the branching and annihilation terms are small perturbations. To exploit the expected time-scale separation,  we perform a canonical transformation from $\xb$ and $\pb$ to $q_i,Q$ and $P_{i},P_Q$, where $i=2,3,...,N$:
\begin{eqnarray}
\label{eq:can}
&&Q = X =\sum_j x_j, \; \; \; q_i=x_i, \nonumber \\
&&p_1=P_Q, \; \; \; p_{i}=P_Q+P_{i}.\nonumber
\end{eqnarray}
This canonical transformation is motivated by our expectation that, at small $\varepsilon$, the total population size $Q$ and the corresponding conjugate momentum $P_Q$ are slowly varying quantities.

In the new variables, the Hamiltonian acquires the form
\begin{eqnarray}
\tilde{H}(\{q_i\},Q, \{p_{q_i}\},P_Q) &=& \tilde{H}_0(\{q_i\},Q, \{p_{q_i}\})  \nonumber \\
&+&\varepsilon \tilde{H}_1(\{q_i\},Q, \{p_{q_i}\},P_Q) \label{eq:H00}
\end{eqnarray}
for $i=2,3,...,N$, where
\begin{eqnarray}
\tilde{H}_0&=&\frac{1}{2}\sum_{i\neq 1} \sum_{j\neq 1, j\in {\cal I}_i}M_{ij}\left[ q_i\left(e^{P_j-P_i}-1\right)+ q_j\left(e^{P_i-P_j}-1\right)\right] \nonumber \\
&+&\sum_{j \in {\cal I}_1}M_{1j}\left[\left(Q-\sum_{i=2}^N q_i\right)\left(e^{P_j}-1\right)+q_j\left(e^{-P_j}-1\right)\right], \nonumber \\
\label{eq:H0} \\
\tilde{H}_1&=&\sum_{i=2}^N q_i\left(e^{P_Q+P_{i}}-1\right)+\frac{q_i^2}{2\kappa_i} \left[e^{-2(P_Q+P_{i})}-1\right]\nonumber \\
&+& \left(Q-\sum_{i=2}^N q_i\right)\left(e^{P_Q}-1\right)\nonumber \\
&+&\frac{\left(Q-\sum_{i=2}^N q_i\right)^2}{2\kappa_1} \left(e^{-2P_Q}-1\right). \label{eq:ham2}
\end{eqnarray}
Notably, the migration Hamiltonian $\tilde{H}_0(\{q_i\},Q, \{p_{q_i}\})$  does not depend on $P_Q$. This reflects the simple fact that, in the absence of branchings and annihilations, the total population size $Q$ is conserved. For brevity we shall omit tildes in $\tilde{H}$,$\tilde{H}_0$ and $\tilde{H}_1$ in the following.

The Hamiltonian equations of motion associated with (\ref{eq:H00})  are:
\begin{eqnarray}
\!\!\!\!\!\!\dot{q}_i&=&\partial_{P_i}{H}_0(\qb,Q, \textbf{P})+\varepsilon \partial_{P_i} {H}_1(\qb,Q, \textbf{P},P_Q),\label{eq:mot1}\\
\!\!\!\!\!\!\dot{P}_{i}&=&-\partial_{q_i}{H}_0(\qb,Q, \textbf{P})-\varepsilon \partial_{q_i} H_1(\qb,Q, \textbf{P},P_Q), \label{eq:mot2} \\
\!\!\!\!\!\!\dot{Q}&=&\varepsilon \partial_{P} H_1(\qb,Q, \textbf{P},P_Q),\label{eq:mot3} \\
\!\!\!\!\!\!\dot{P_Q}&=&-\partial_Q {H}_0(\qb,Q, \textbf{P})-\varepsilon \partial_Q H_1(\qb,Q, \textbf{P},P_Q),\label{eq:mot4}
\end{eqnarray}
where $\qb=(q_2,q_3,...,q_N)$ and $\textbf{P}=(P_2,P_3,...,P_N)$. We note that no approximations have been made so far, and Eqs.~(\ref{eq:mot1})-(\ref{eq:mot2}) are valid for any $\varepsilon$.

The perturbation scheme that we have developed for Eqs.~(\ref{eq:mot1})-(\ref{eq:mot4}) assumes the following time-scale separation scenario. At $\varepsilon \ll 1$ the total population size $Q$ varies in time on a long time scale of order $\varepsilon^{-1} \gg 1$.  The local population sizes $q_i$ quickly (on a time scale of order unity) adjust to the instantaneous value of $Q$. Moreover, at given $Q$, the dynamics of the local population sizes are essentially deterministic. Having this scenario in mind, we can look for the solution as $q_i=\tilde{q}_i(Q)+\varepsilon q_i^{(1)}$ and $P_i=\varepsilon P_i^{(1)}$, where $\tilde{q}_i(Q)$ is the fixed point of Eq.~(\ref{eq:mot1}) with $\varepsilon=0$ \emph{for an instantaneous value of} $Q$, see Eq.~(\ref{eq:motion2a}) below. The smallness of $P_i$ reflects the expectation that the fast variables evolve almost deterministically. In this case   $H_0$ in Eq.~(\ref{eq:H0}) can be linearized with respect to $P_i$, and we obtain:
\begin{eqnarray}
&&{H}_0(\qb,Q, \textbf{P})=\varepsilon \textbf{P}^{(1)} \cdot \textbf{h}(\qb,Q)+O(\varepsilon^2), \label{eq:smallph}   \\
&&h_k=  \sum_{j\neq 1, j\in {\cal I}_k}M_{kj}\left( q_j-q_k\right)    \label{eq:h}  \\
&&+\sum_{j \in {\cal I}_1}M_{1j}\delta_{kj} \left(Q -q_k -\sum_{i=2}^N q_i  \right),\;\;\;k=2,3,\dots, N, \nonumber
\end{eqnarray}
where $\delta_{kj}$ is the Kronecker delta. It is seen from Eq.~(\ref{eq:smallph}) that $H_0$ is actually of the first order in $\varepsilon$, rather than of the zeroth order as could have been na\"{\i}vely expected from Eq.~(\ref{eq:H00}).
In the leading order in $\varepsilon$   Eqs.~(\ref{eq:mot1}) and (\ref{eq:mot2}) become
\begin{eqnarray}
0&=&h_i(\tilde{\qb},Q) \label{eq:motion2a}, \\
0&=&\partial_{q_i} \textbf{P}^{(1)} \cdot \textbf{h}(\qb,Q)+\partial_{q_i} H_1(\qb,Q, \textbf{0},P_Q). \label{eq:motion2b}
\end{eqnarray}
Solving the set of algebraic equations~(\ref{eq:motion2a}) for $\tilde{q}_i$, and using
Eqs.~(\ref{eq:smallph}) and (\ref{eq:h}), we obtain
\begin{eqnarray}
&&\tilde{q}_i(Q)=Q/N, \ \ i=2,3,...,N,\label{eq:tildq}
\end{eqnarray}
in agreement with our mean-field results in the fast-migration limit, see Sec. 1.

Now let us take the total derivative of Eq.~(\ref{eq:motion2a}) with respect to $Q$:
\begin{eqnarray}
\label{eq:motion3}
\frac{\partial \textbf{P}^{(1)} \cdot \textbf{h}(\qb,Q)}{\partial \tilde{q}_j}\frac{d \tilde{q}_j}{d Q} +\frac{\partial \textbf{P}^{(1)} \cdot \textbf{h}(\qb,Q)}{\partial Q }=0,
\end{eqnarray}
and combine this result with Eq.~(\ref{eq:motion2b}) multiplied by $d\tilde{q}_i/dQ$ and summed over $i$ from $2$ to $N$.
We obtain
\begin{eqnarray}
\label{eq:motion5}
\frac{\partial H_1(\tilde{\qb},Q, \textbf{0},P_Q)}{\partial{\tilde{q}_j}}\frac{d \tilde{q}_j}{d Q}=\frac{\partial \textbf{P}^{(1)} \cdot \textbf{h}(\qb,Q)}{\partial Q},
\end{eqnarray}
where the summation convention is used. Using Eqs.~(\ref{eq:smallph} ) and (\ref{eq:motion5}) in Eq.~(\ref{eq:mot4}), and restoring the original time, we can rewrite  Eqs.~(\ref{eq:mot3}) and (\ref{eq:mot4}) as
\begin{eqnarray}
\label{eq:motion6}
\dot{Q}&=& \partial_{P} H_1(\tilde{\qb}(Q),Q, \textbf{0},P_Q)=\partial_{P} H_{slow}(Q,P_Q), \\
\dot{P}&=&-\partial_{q_j} H_1(\tilde{\qb}(Q),Q, \textbf{0},P_Q)\frac{d \tilde{q}_j}{d Q}-\partial_Q H_1(\tilde{\qb}(Q),Q \textbf{0},P_Q),\nonumber \\
&=&-\partial_Q H_{slow}(Q,P_Q),
\end{eqnarray}
where we have introduced the slow Hamiltonian
\begin{eqnarray}
\label{eq:hamslow1}
H_{slow}\left(Q,P\right)&=&H_1(\tilde{\qb}(Q),Q, \textbf{0},P_Q)\nonumber \\
&=& \sum_{i=2}^N \left\{\frac{Q}{N}\left(e^{P_Q}-1\right)+\frac{Q^2}{2N^2 \kappa_i} \left(e^{-2 P_Q}-1\right)\right\}\nonumber \\
&+& \frac{Q}{N}\left(e^{P_Q}-1\right)+\frac{Q^2}{2N^2 \kappa_1} \left(e^{-2P_Q}-1\right) \nonumber \\
&=&Q \left(e^{P_Q}-1\right)+\frac{Q^2}{2 \bar{\kappa} } \left(e^{-2P_Q}-1\right),
\end{eqnarray}
where $\bar{\kappa}$ is defined in Eq.~(\ref{eq:mf2}). This concludes our derivation of the effective one-population Hamiltonian [see Eq.~(18) of the main text] which describes slow large fluctuations of the total meta-population size.


\begin{thebibliography}{99}
\bibitem{Levins}  R. Levins, Bull. Entomol. Soc. Amer. \textbf{15}, 237 (1969).
\bibitem{Hanski} I. Hanski, \emph{Metapopulation Ecology} (Oxford University Press, Oxford, 1999).
\bibitem{Hanski2} \emph{Ecology, Genetics, and
Evolution in Metapopulations}, edited by I. Hanski  and O. Gaggiotti (Elsevier Academic Press, Burlington, 2004).
\bibitem{endangered} R. Bierregaard, T. Lovejoy, V. Kapos, A. Dossantos, and R.W. Hutchings, Bioscience \textbf{42}, 859 (1992);
J. F. Quinn and A. Hastings,  Conserv. Biol. \textbf{1}, 198 (1987);
S. K. Robinson,  F. R. Thompson, T. M. Donavon,   D. R. Whitehead, and J. Faaborg, Science \textbf{267}, 1987 (1995);
I. Turner, K. Chua, J. Ong, B. Soong, and H. Tan, Conserv. Biol. \textbf{10}, 1229 (1996).
\bibitem{corridor} D. Simberloff and J. Cox, Conserv. Biol. \textbf{1}, 63 (1987).
\bibitem{Ellner} S. P. Ellner, E. McCauley, B. E. Kendall, C. J. Briggs, P. R. Hosseini, S. N. Wood, A. Janssen,
M. W. Sabelis, P. Turchin, R. M. Nisbet, and W. W. Murdoch, Nature \textbf{412} 538 (2001).
\bibitem{Molofsky} J. Molofsky and J. Ferdy, Proc. Natl. Acad. Sci. USA \textbf{102}, 3726 (2005).
\bibitem{Holoyak} M. Holyoak and S. P. Lawler, Ecology \textbf{77}, 1867 (1996).
\bibitem{Dey} S. Dey and A. Joshi, Science \textbf{312}, 434 (2006).
\bibitem{Kerr} B. Kerr, C. Neuhauser,  B. J. M. Bohannan, and A. M. Dean, Nature \textbf{442}, 75 (2006).
\bibitem{Shnerb} G. Yaari, Y. Ben-Zion, N. M. Shnerb, and D. A. Vasseur,  Ecology \textbf{93}, 1214 (2012). 
\bibitem{Gurney}  W. S. C. Gurney and R. M. Nisbet, Am. Nat. \textbf{112}, 1075 (1978).
\bibitem{McKane} D. Alonso and A. McKane, Bull. Math. Biol. \textbf{64}, 913 (2002).
\bibitem{Ross} J.V. Ross, J. Math. Biol. \textbf{52}, 788–806 (2006); Bull. Math. Biol. \textbf{68}, 417 (2006).
\bibitem{TREE} O. Ovaskainen and B. Meerson, Trends in Ecology and Evolution \textbf{25}, 643 (2010).
\bibitem{turner} J. W. Turner and M. Malek-Mansour, Physica A 93, 517 (1978).
\bibitem{ElgartPRE70} V. Elgart and A. Kamenev, Phys. Rev. E \textbf{70}, 041106 (2004).
\bibitem{Kessler} D. A. Kessler and N. M. Shnerb, J. Stat. Phys. 127, 861 (2007).
\bibitem{AM1} M. Assaf and B. Meerson, Phys. Rev. E \textbf{75}, 031122 (2007).
\bibitem{suppl} See Appendix.
\bibitem{comparison} Although for equal carrying capacities $T_{\mu \rightarrow \infty}$ is exponentially
larger than $T_{\mu=0}$, for arbitrary values of $\kappa_i$ the ratio  $T_{\mu=0}/T_{\mu \rightarrow \infty}$ can be arbitrary.
\bibitem{AM2006} M. Assaf and B. Meerson, Phys. Rev. Lett. \textbf{97}, 200602 (2006).
\bibitem{AssafPRE81} M. Assaf and B. Meerson, Phys. Rev. E \textbf{81}, 021116 (2010).
\bibitem{MS11a} B. Meerson and P.V. Sasorov, Phys. Rev. E \textbf{83}, 011129  (2011).
\bibitem{Kubo} R. Kubo, K. Matsuo, and K. Kitahara, J. Stat. Phys. \textbf{9}, 51  (1973).
\bibitem{Hu} G. Hu, Phys. Rev. A \textbf{36}, 5782 (1987).
\bibitem{Peters} C.S. Peters, M. Mangel, and R. F. Costantino,  Bull. Math. Biol. \textbf{51}, 625 (1989).
\bibitem{DykmanPRE100} M. I. Dykman, E. Mori, J. Ross, and P. M. Hunt, J. Chem. Phys. \textbf{100}, 5735 (1994).
\bibitem{DykmanPRL101} M. I. Dykman, I. B. Schwartz, and A. S. Landsman, Phys. Rev. Lett. \textbf{101}, 078101 (2008).
\bibitem{MeersonPRE77} A. Kamenev and B. Meerson, Phys. Rev. E \textbf{77}, 061107 (2008).
\bibitem{KamMS} A. Kamenev, B. Meerson, and B. Shklovskii, Phys. Rev. Lett. \textbf{101}, 268103 (2008).
\bibitem{AKMmod} M. Assaf, A. Kamenev, and B. Meerson, Phys. Rev. E \textbf{78}, 041123 (2008).
\bibitem{MeersonPRE79} M. Assaf, A. Kamenev, and B. Meerson, Phys. Rev. E \textbf{79}, 011127 (2009).
\bibitem{KhasinPRL103} M. Khasin and M. I. Dykman, Phys. Rev. Lett. \textbf{103}, 068101 (2009).
\bibitem{MeersonPRE81} M. Khasin, M. I. Dykman, and B. Meerson, Phys. Rev. E \textbf{81}, 051925 (2010).
\bibitem{AMS} M. Assaf, B. Meerson, and P. V. Sasorov, J. Stat. Mech. P07018 (2010).
\bibitem{LM2011} I. Lohmar and B. Meerson, Phys. Rev. E \textbf{84}, 051901 (2011).
\bibitem{GM} O. Gottesman and B. Meerson, Phys. Rev. E \textbf{85}, 021140 (2012).
\bibitem{Doering} C.R. Doering, K.V. Sargsyan, and L.M. Sander, Multiscale Model. and Simul. \textbf{3},
283 (2005).
\bibitem{infinity} That $p_x=p_y=-\infty$ at the extinction
fixed point stems from the absence of linear in $m$ and $n$ death processes in this model. This divergence causes no harm,
as the action along the instanton is finite \cite{ElgartPRE70,Kessler}.
\bibitem{Stepanov} A. I. Chernykh and M. G. Stepanov, Phys. Rev. E \textbf{64}, 026306
(2001).
\bibitem{degenerate} For $(T_{\mu=0})^{-1} \ll \mu K \ll 1$ the MTE can be found analytically from a degenerate perturbation theory applied directly to the master equation (\ref{eq:master_master}). We did this calculation and arrived at the same $T$ as given by Eq.~(\ref{S1}), as expected on heuristic grounds.
\bibitem{book} R. Rajaraman, \textit{Solitons and Instantons} (Amsterdam, North Holland, 1987).
\end{thebibliography}
\end{document}